\documentclass[%
 aip,
 amsmath,amssymb, xurl,
 reprint,%
]{revtex4-1}
\usepackage{siunitx}
\usepackage{graphicx}
\usepackage{dcolumn}
\usepackage{bm}
\pdfoutput=1 
\usepackage[utf8]{inputenc}
\usepackage[T1]{fontenc}
\usepackage{mathptmx, float}
\usepackage{etoolbox}
\PassOptionsToPackage{hyphens}{url}\usepackage{hyperref}

\usepackage{easyReview}
\definecolor{mypurple}{HTML}{800080}

\newcommand{\sbonlinecite}[1]{[\onlinecite{#1}]}

\makeatletter
\def\@email#1#2{%
 \endgroup
 \patchcmd{\titleblock@produce}
  {\frontmatter@RRAPformat}
  {\frontmatter@RRAPformat{\produce@RRAP{*#1\href{mailto:#2}{#2}}}\frontmatter@RRAPformat}
  {}{}
}%
\makeatother

\begin{document}

\preprint{AIP/123-QED}

\title{Designing high-performance superconductors with nanoparticle inclusions: comparisons to strong pinning theory}

\author{Sarah C. Jones}
\affiliation{Department of Physics, Colorado School of Mines, Golden, Colorado 80401, USA}

\author{Masashi Miura}
\affiliation{Graduate School of Science and Technology, Seikei University, Tokyo 180-8633, Japan}

\author{Ryuji Yoshida}
\affiliation{Nanostructures Research Laboratory, Japan Fine Ceramics Center, Nagoya 456-8587, Japan}

\author{Takeharu Kato}
\affiliation{Nanostructures Research Laboratory, Japan Fine Ceramics Center, Nagoya 456-8587, Japan}

\author{Leonardo Civale}
\affiliation{Materials Physics and Applications Division, Los Alamos National Laboratory, Los Alamos, NM 87545, USA}

\author{Roland Willa}
\affiliation{Institute for Theory of Condensed Matter, Karlsruhe Institute of Technology, 76131 Karlsruhe, Germany}
\affiliation{Heidelberger Akademie der Wissenschaften, 69117 Heidelberg, Germany}

\author{Serena Eley}
\affiliation{Department of Physics, Colorado School of Mines, Golden, Colorado 80401, USA}
\email[Corresponding author: ]{serenaeley@mines.edu}

\date{\today}

\begin{abstract}
One of the most promising routes for achieving unprecedentedly high critical currents in superconductors is to incorporate dispersed, non-superconducting nanoparticles to control the dissipative motion of vortices.  However, these inclusions reduce the overall superconducting volume and can strain the interlaying superconducting matrix, which can detrimentally reduce $T_c$. Consequently, an optimal balance must be achieved between the nanoparticle density $n_p$ and size $d$. Determining this balance requires garnering a better understanding of vortex-nanoparticle interactions, described by strong pinning theory.  Here, we map the dependence of the critical current on nanoparticle size and density in (Y$_{0.77}$,Gd$_{0.23}$)Ba$_2$Cu$_3$O$_{7-\delta}$ films in magnetic fields up to \SI{35}{\tesla}, and compare the trends to recent results from time-dependent Ginzburg-Landau simulations. We identify consistencies between the field-dependent critical current $J_c(B)$ and expectations from strong pinning theory. Specifically, we find that that $J_c \propto B^{-\alpha}$, where $\alpha$ decreases from $0.66$ to $0.2$ with increasing density of nanoparticles and increases roughly linearly with nanoparticle size $d/\xi$ (normalized to the coherence length). At high fields, the critical current decays faster ($\sim B^{-1}$), suggestive that each nanoparticle has captured a vortex.  When nanoparticles capture more than one vortex, a small, high-field peak is expected in $J_c(B)$. Due to a spread in defect sizes, this novel peak effect remains unresolved here. Lastly, we reveal that the dependence of the vortex creep rate $S$ on nanoparticle size and density roughly mirrors that of $\alpha$, and compare our results to low-$T$ nonlinearities in $S(T)$ that are predicted by strong pinning theory.

\end{abstract}

\maketitle

\section{Introduction}\label{sec:intro}

Recently, the demand for high-performance superconductors --- having large current-carrying capacities and resiliency to strong magnetic fields --- has increased. These materials are necessary constituents in a multitude of applications,\cite{Hahn2019, Molodyk2021} including next-generation fusion reactors for clean energy production,\cite{Creely2020, greenwald2020, Whyte2016, Molodyk2021, APS2018} cables for power transmission, particle accelerators for high-energy physics,\cite{Rossi2019, instruments3040062} magnetic resonance imaging for radiology,\cite{Parizh2016, Nowogrodzki2018, Moser2017, global2021} and nuclear magnetic resonance spectroscopy for biological applications.\cite{MAEDA201980, Moser2017} Meeting established metrics for operating conditions requires careful design of the conductor material, necessitating substantial changes from the intrinsic crystalline structure.

Incorporating nanometer-size inclusions into the microstructure of high-$T_c$ copper- and iron-based superconductors can drastically increase the material's critical current density $J_c$, below which transport is dissipation-free.\cite{Eley2021, Macmanus-Driscoll2004, Gurevich2011b, Kwok2016, Wimbush_2010, Ercolano_2011, Harrington_2008, Miura2017, Miura2016, Miura2013k, Miura2011b, Haugan2003, Haugan2004, Selvamanickam2015} This is because nanoparticles arrest dissipative vortex motion. In fact, while current-induced Lorentz forces and thermal energy tend to propel vortices, lines of quantized magnetic flux $\Phi_0$ piercing the superconductor even in minute magnetic fields, defects can provide efficient pinning forces to counteract this motion. 

Interest in nanoparticle inclusions spans beyond their ability to boost $J_c$, but also encompasses accompanying increases in the irreversibility field\cite{Miura2016} $H_{irr}$ and decreases in the rate of thermally activated vortex motion (creep).\cite{Haberkorn2012d, Rouco2014d, Miura2011b, Miura2013k}  Most notably, $\mathrm{Y}\mathrm{Ba_2}\mathrm{Cu_3}\mathrm{O_{7-\delta}}\mathrm{ (YBCO)}$ films containing nanoparticles exhibit a non-monotonic temperature-dependent creep rate $S(T)$, dipping to very low values at $T \!\sim\! 0.3 T_c \text{--} 0.5 T_c$, which has not otherwise been achieved in $\mathrm{YBCO}$ single crystals\cite{Eley2017a} nor iron-based films containing nanoparticles. Understanding the mechanism behind this dip is key to further minimizing $S$ while maximizing $J_c$ in $\mathrm{YBCO}$ and other superconductors. Remarkably, the beneficial effects of nanoparticles cannot be presumed.  Their effects depend specifically on the intrinsic microstructure and measurement conditions (e.g., temperature $T$ and applied magnetic field $H$). For example, $\mathrm{Ba}\mathrm{Zr}\mathrm{O}_{3}$ inclusions unfavorably increased the creep rate $S$ in $\mathrm{Ba}\mathrm{Fe}_2(\mathrm{As}_{0.67}\mathrm{P}_{0.33})_2$ thin films exposed to low magnetic fields.\cite{Eley2017}

Despite the vast studies demonstrating remarkable advances in growth techniques and accompanying enhancements of $J_c$,\cite{Macmanus-Driscoll2004, Gurevich2011b, Kwok2016, 
Wimbush_2010, Ercolano_2011, Harrington_2008, Miura2017, Miura2016, Miura2013k, Miura2011b, Haugan2003, Haugan2004} many open questions remain largely unanswered: Why do nanoparticles induce non-monotonic temperature-dependent creep rates in $\mathrm{YBCO}$ films and what controls the local minimum in $S(T) $? Under what conditions do nanoparticles act collectively, versus as strong pinning centers? What is the maximum non-superconducting volume fraction that can be consumed by nanoparticle inclusions to maximize $J_c$ without compromising the critical temperature $T_c$?  Furthermore, what is the optimal combination of nanoparticle size $d$ and density $n_p$ to constitute this non-superconducting volume? Answering these questions will enable optimized design of the microstructure in  superconductors.

Here, we present a comparative study of vortex pinning in (Y$_{0.77}$,Gd$_{0.23}$)Ba$_2$Cu$_3$O$_{7-\delta}$ films containing different sizes and densities of nanoparticle inclusions and compare the results to the expectations of strong pinning theory.\cite{Willa2017, Willa2018b, Buchacek2018, Buchacek2019}
A previous study\cite{Miura2013k} mapped the dependence of nanoparticle size and density on $J_c$ at high temperatures $T/T_c \sim 0.7\text{-}0.9$.  We build upon this work by presenting a similarly systematic study at significantly lower temperatures $T/T_c \sim 0.05\text{-}0.5$ as well as at high fields up to \SI{35}{\tesla}.  Mapping the trends at low temperatures enables direct comparisons to strong pinning theory formalisms that do not consider thermal activation, whereas high field measurements explore the high vortex density regime.  We present four main results: First, at low fields, we find that $J_c$ increases roughly linearly with nanoparticle density $n_p$, then $J_c(n_p)$ flattens out at higher densities. Second, we observe the two-stage power-law behavior of the critical current $J_c \!\propto\! B^{-\alpha}$ that is expected, and previously observed,~\cite{Eley2017a,VanderBeek2002a, Miura2011b,Haberkorn2012d, Haberkorn2015d, Maiorov2009a, Ijaduola2012, Senatore2016, Coll2013, Long2005} for the strong-pinning scenario. We subsequently find that $\alpha$ decreases with increasing nanoparticle density and increases with nanoparticle diameter $d/\xi$, normalized to the coherence length $\xi \equiv \xi_{ab}$. Third, we study $J_c(B)$ at high vortex densities to look for the theoretically predicted peak effect that may occur when individual nanoparticles capture more than one vortex.  Last, we find that samples containing denser nanoparticles produce slower creep rates, whereas larger nanoparticle diameters $d/\xi$ result in increased creep rates at low temperatures $T \ll T_c$.

\section{Theoretical Background}\label{sec:theory}

The dynamics of vortices depends on the interplay between current-induced forces, thermal energy, vortex elasticity, and pinning forces exerted by defects. Broadly, the complexity of the problem is dictated by three constituents: First, the superconductor's intrinsic properties defining the 'pure' vortex state, i.e. penetration depth $\lambda$, coherence length, and electronic mass anisotropy $\gamma^2=m_c/m_a$. Second, artificial or naturally occurring inhomogeneities in the superconducting matrix provide a defect structure (pinning landscape) that impedes vortex motion for sufficiently small driving forces. The third contribution originates from the sample geometry that is known to define a panoply of geometric and surface effects.\cite{Bean1964, Zeldov1994, Benkraouda1996, Benkraouda1998, Brandt1999a, Mikitik2004, Willa2014}

Segments of a vortex line or vortex bundles can be pinned by the independent action of large defects (strong pinning) or by the collective action of many weak, small pins (weak pinning).\cite{Larkin1979, Blatter1994, Blatter2004a, Ovchinnikov1991} In the absence of thermal effects, the optimal pinning configuration results from balancing the energy cost to deform a vortex state away from its lattice arrangement by an energy gain when vortices accommodate to neighboring pinning sites. Along a vortex line, segments of characteristic length $L_c$ are independently pinned, while in the transverse direction, pinning of vortices separated by more than $R_c$ is uncorrelated. The correlation volume thus becomes $V_c \propto L_c R_c^2$; its parametric dependence has been calculated for the weak and strong pinning cases.\cite{Larkin1979, Blatter1994, Blatter2004a, Ovchinnikov1991} 
Though pinning disrupts long-range order of the vortex lattice, short-range order exists within $V_c$, which can contain a single vortex or a bundle of vortices, depending on the vortex density, where the vortex spacing is $a_0 \!=\! (4/3)^{1/4}(\Phi_0/\mu_0 H)^{1/2}$.

At finite temperatures, these single vortices or bundles hop out of their pinning sites due to thermal activation, such that the vortex arrangement relaxes (creeps) from its out-of-equilibrium arrangement.
This is often captured by measuring the relaxation rate of the persistent currents,\cite{Yeshurun1996} creep-induced curvature in $IV$ characteristics\cite{Kwok2016, Buchacek2019} near $I\gtrsim I_c$ for $V\sim I^n$, or imaging studies (e.g. scanning tunneling microscopy).\cite{Willa2020a}  Moreover, at sufficiently high temperatures, the vortex state undergoes a melting transition in which pinning becomes ineffective.\cite{Zeldov1995, Li2002, Li2004, Koshelev2019, Hardy2020}

\subsection{Strong pinning theory: predictions for the critical current }

The limit of dilute, strong defects has been studied analytically in recent years. Contrary to the qualitative scaling arguments of weak collective pinning theory, the strong pinning framework provides quantitative results for various observables ($J_{c}$ \cite{Labusch1969, Larkin1979, Blatter2004a, Koshelev2011, Willa2017}, Campbell penetration \cite{Willa2015a, Willa2015b, Willa2016, Willa2018c}, zero- and finite-temperature current-voltage characteristics.\cite{Thomann2012, Thomann2017, Buchacek2018, Buchacek2019}) Specifically, the critical current has been shown\cite{Blatter2004a} to follow $J_{c} = J_{\mathrm{dp}} (n_{p} a_{0} \xi^{2}) (f_{p}/\varepsilon_{0})^{2}$, 
over a wide field range, with $J_{\mathrm{dp}}$ being the depairing current, $n_{p}$ the defect density, $f_{p}$ the elementary force provided by a single defect, and $\varepsilon_0$ the vortex line energy. This result implies a power-law field dependence $J_c \propto B^{-\alpha}$ with $\alpha = 1/2$, in agreement with an early work by Larkin and Ovchinnikov. \cite{Larkin1979}

A recent study\cite{Willa2017} demonstrated that the effect of defects trapping vortices \emph{transverse} to the direction of the force increases the power $\alpha$ by $1/8$ to $5/8$. Furthermore, numerical simulations of the strong-pinning regime have revealed that the power-law behavior $J_{c} \!\propto\! B^{-\alpha}$ persists beyond the limit of diluted defects, albeit with a reduced power $\alpha < 5/8$. Specifically, it was found that the power $\alpha$ depends primarily on the volume fraction $V$ occupied by the defect, only weakly depends on the defect size, and follows an empiric law $\alpha = 0.026 \text{-} 0.083 \ln(V)$ (for $V > 10^{-3}$), see also Ref.~\sbonlinecite{Lao2019}.

As the magnetic field increases, so does the vortex density, such that vortex-vortex interactions become increasingly relevant and vortices compete for pinning sites. When the vortex density surpasses the number of available pinning sites, some vortices may be effectively immobilized by caging effects (pinned vortices repel interstitial vortices)~\cite{PhysRevB.85.012505, Berdiyorov2006, PhysRevB.74.174512} or pinning centers may host more than one vortex.~\cite{Willa2017}  This change in pinning behavior can evoke a change in the magnetic and transport properties.  For example, Ref.~[\onlinecite{Willa2018b}] predicts a novel type of peak-effect in $J_{c}(B)$ that appears at high fields as the system transitions from nanoparticles capturing one to capturing multiple vortices. Despite rapid improvements in controlling the size of incorporated nanoparticles, experimental evidence for this effect remains elusive. Finally, at large fields, where the inter-vortex distance falls below the typical defect size, the critical current is expected to decay faster $J_{c} \propto B^{-(1+\beta)}$, where $\beta > 0$ accounts for the competition between vortices for the same defect.

\subsection{Strong pinning theory: predictions for vortex creep}

Material disorder creates a landscape of potential energy wells in which vortices localize to reduce their core energies by a pinning energy $U_c$, defined at low currents $J \rightarrow 0$. This energy landscape is then tilted by a current, reducing the energy barrier that a pinned vortex must surmount to escape from the well to $U(J)$.  The approximate time for thermal activation over the barrier is
\begin{equation}
    t=t_0e^{U(J)/k_BT},\label{eq:arrhenius}
\end{equation} where $t_0$ is a microscopic attempt time for vortex creep (typically \SI{1}{\micro\second}).\cite{Yeshurun1996}  

Generally, the creep energy barrier is thought to depend on current as $U(J) \approx U_c(1-J/J_c)^n$, where $n$ is determined by the pinning landscape.  The Anderson-Kim model\cite{Anderson1964, Blatter1994} predicts $n=1$; as it disregards vortex elasticity and vortex-vortex interactions, it is often a useful approximation at low temperatures ($T \ll T_c$), low magnetic fields, and in the early stages of the relaxation process ($J\lesssim J_{c0}$), where $J_{c0}$ is the creep-free critical current. For currents $J \ll J_{c0}$, the weak collective creep paradigm, which considers vortex elasticity, models the energy barrier as $U(J) = U_c(J_{c0}/J)^\mu$, where $\mu$ is the glassy exponent that depends on the size of the vortex bundle that hops during the creep process and its dimensionality.\cite{Blatter1994, Vinokur1995}  Interpolating between the two regimes, it is common to model the energy barrier as $U(J)=(U_c/\mu)[(J_{c0}/J)^\mu-1]$, which consequently covers a broad range of currents. Combining Eq.~\ref{eq:arrhenius} with this interpolation formula, we expect $J_c$ to decay over time as 
\begin{equation}
J(t) = J_{c0}\left[1+\frac{\mu k_BT}{U_c}\ln(t/t_0)\right]^{-1/\mu}\label{eq:Jt}.
\end{equation} The vortex creep rate then parameterizes this decay as
\begin{equation}
    S \equiv \left|\frac{d \ln J}{d \ln t}\right| \label{eq:Sdefinition}
\end{equation}
such that, in the collective pinning scenario, the creep rate depends on the temperature and $U_c$ as
\begin{equation}
    S = \frac{k_BT}{U_c+\mu k_B T \ln(t/t_0)}.\label{eq:ST}
\end{equation}
Recent results from strong pinning theory\cite{Buchacek2019}, however, predict that $n \approx 3/2$, resulting in
\begin{equation}
    U(J) \approx U_c\left(1-J/J_c\right)^{3/2}
\end{equation}
for $J<J_c$.  This leads, instead, to a thermal creep rate of\cite{Buchacek2019}
\begin{equation}
    S = \frac{2}{3}\frac{(k_BT/U_c)^{2/3}[\ln(t/t_0)]^{-1/3}}{1-[(k_BT/U_c)\ln(t/t_0)]^{2/3}}.\label{eq:creepstrongpinning}
\end{equation}.

\section{Samples and Experimental Procedure}

In this work, we studied one reference film of $\mathrm{(Y,Gd)Ba}_2\mathrm{Cu}_3\mathrm{O}_{7-x}$ and seven films containing $\mathrm{Ba}M\mathrm{O}_3$ inclusions (for $M = \mathrm{Zr}$, $\mathrm{Hf}$, or $\mathrm{Sn}$). The reference film was 900~nm thick and the thicknesses of the films containing $\mathrm{Ba}M\mathrm{O}_{3}$ nanoparticles varied from 290~nm to 910~nm. Table \ref{tab:samples} summarizes the parameters of each sample and assigns a sample ID that is referenced in the data. All films were grown epitaxially on buffered Hastelloy substrates by means of the trifluoroacetate metal-organic deposition process from $\mathrm{Y}$-, $\mathrm{Gd}$-, and $\mathrm{Ba}$-trifluoroacetates and $\mathrm{Cu}$-naphthenate solutions with the cation ratio of 0.77:0.23:1.5:3. For the films containing $\mathrm{Ba}M\mathrm{O}_{3}$ nanoparticles, $\mathrm{Zr}$-, $\mathrm{Hf}$-, or $\mathrm{Sn}$-naphthenate was introduced into the precursor solution. The interposing buffer was a stack of $\mathrm{Ni}\mathrm{Cr}\mathrm{Fe}\mathrm{O}$, $\mathrm{Gd}_2\mathrm{Zr}_2\mathrm{O}_{7}$, $\mathrm{Y}_2\mathrm{O}_{3}$, $\mathrm{Mg}\mathrm{O}$ (deposited using ion beam assisted deposition), $\mathrm{La}\mathrm{Mn}\mathrm{O}_{3}$, and $\mathrm{Ce}\mathrm{O}_{2}$ layers.  Note that all samples contain a sparse distribution ($0.3 \times 10^{21}/m^3$) of $\mathrm{Y}_2\mathrm{Cu}_2\mathrm{O}_5$ (225) nanoparticle precipitates that naturally form during the growth process. Further details of the growth process are elaborated elsewhere.\cite{Miura2017} 
\begin{table}[h!] 
\caption{Sample characteristics, including the primary nanoparticle type and volume fraction occupied by the nanoparticles.  *In the reference sample ({\#}7) the dominant defects are $\mathrm{Y}_2 \mathrm{Cu}_2 \mathrm{O}_5$ (225) precipitates, for which the density and diameter is given. All other samples contain a similar density and size of 225 precipitates.}\label{tab:samples}
\begin{ruledtabular}
\begin{tabular}{cccccc}
    sample & film  & Nanoparticle & density $n_p$ & mean & volume\\[-.2em]
    ID  & thickness $h$ & (NP) & $\times 10^{21}$ & diameter & fraction \\[-.2em]
    & [nm] & type & [NP/m$^3$]  & $d$ [nm] & [\%]\\
    \hline
    1 & 380 & BaZrO$_3$ & 5.0 & 28 & 5.6 \\
    2 & 380 & BaZrO$_3$ & 8.0 & 28 & 8.7 \\
    3 & 290 & BaZrO$_3$ & 20  & 18 & 5.9 \\
    4 & 290 & BaZrO$_3$ & 30 & 18 & 8.7 \\
    5 & 290 & BaHfO$_3$ & 100 & 12 & 8.6 \\
    6 & 900 & BaHfO$_3$ & 65  & 14 & 8.9 \\
    7 & 900 & --* & 0.03 & 94 & 1.3*\\ 
    9 & 910 & BaSnO$_3$ & 2.0 & 40 & 6.5 \\ 
\end{tabular}
\end{ruledtabular}
\end{table}

\begin{figure}
    \begin{center}
    \includegraphics[width=1\linewidth]{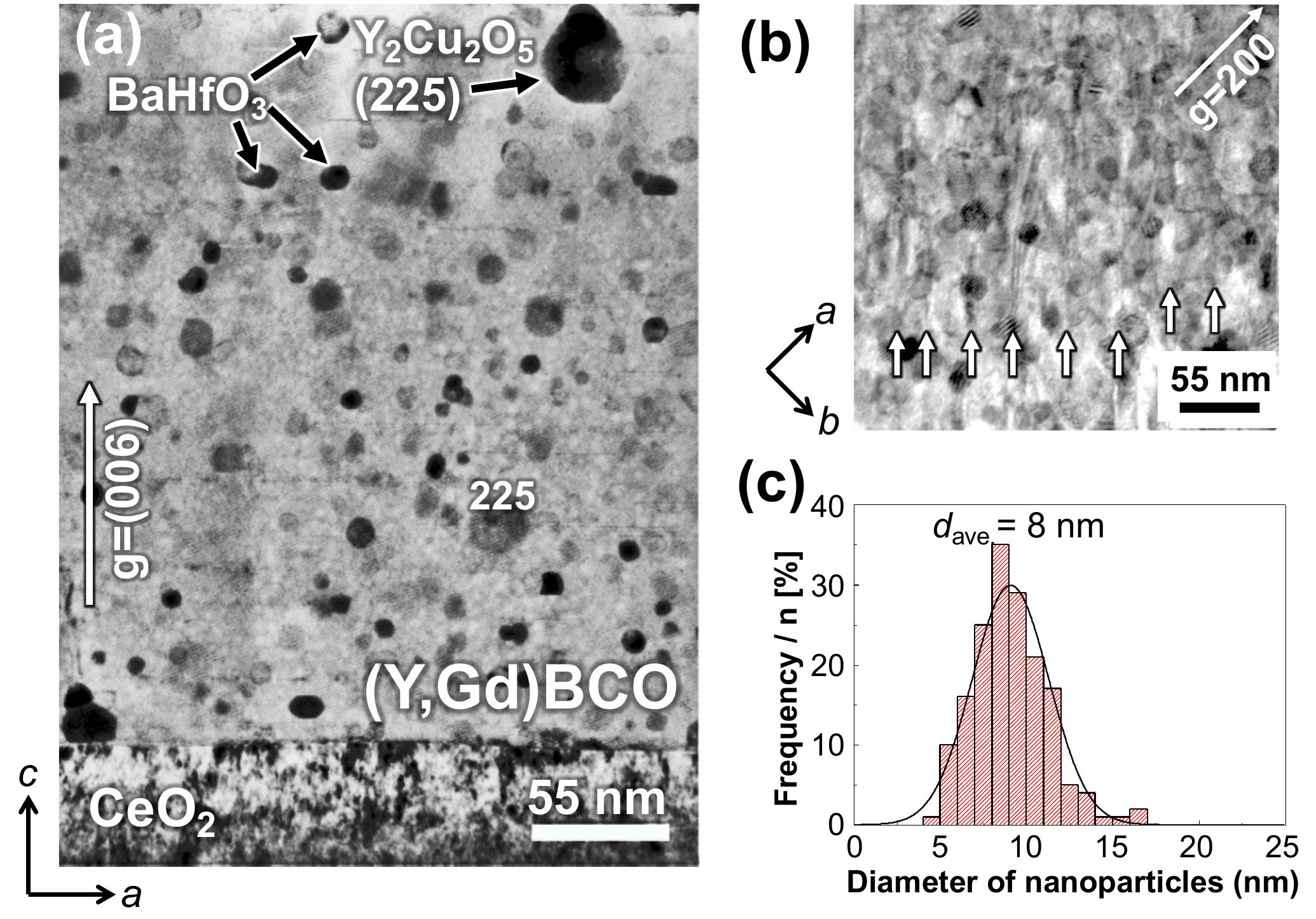}

    \caption{(a) Cross-sectional and (b) planar view transmission electron micrographs of the \SI{290}{\nano\meter} thickness (Y,Gd)BCO film containing 9 vol.\% of BaHfO$_3$ nanoparticles. The film contains a high density of randomly dispersed BaHfO$_3$ nanoparticles and a high density of twin boundaries (white arrows). (c) Histogram showing distribution of particle sizes, fitted to a Gaussian distribution (black curve).}\label{fig:TEM}
    \end{center}
\end{figure}

To map the average size and density of nanoparticle inclusions, the films were characterized via transmission electron microscopy (TEM) and energy dispersive spectroscopy.\cite{Miura2013k} For example, Fig. \ref{fig:TEM}(a,b) shows TEM images of a (Y,Gd)BCO film containing, on average, \SI{8}{\nano\meter}-sized BaHfO$_3$ nanoparticles that was processed under the same conditions as films in this study. As shown in the histogram in Fig. \ref{fig:TEM}(c), all films contain a finite distribution of nanoparticle sizes and Table \ref{tab:samples} indicates the {\it mean} diameter. Note that this study benefits from films grown using a modified process, described in Ref.~[\onlinecite{Miura2017}], that results in a relatively narrow distribution range.

Incorporation of Ba$M$O$_3$ nanoparticles in these films is accompanied by an increased density of two types of planar defects: c-axis oriented twin boundaries and stacking faults along the ab-plane. TEM studies showed that the stacking faults are short ($50\text{-}\SI{100}{\nano\meter}$) and do not segment the twin boundaries, which maintain their integrity throughout the film thickness (contrary to the films studied in Ref.~[\onlinecite{Rouco2014a}]).  The twin boundary spacing ranges from \textasciitilde $\SI{25}{\nano\meter}$ in the film containing the highest density of nanoparticles to \textasciitilde$\SI{45}{\nano\meter}$ in the reference film. For more details, see Ref.~[\onlinecite{Miura2016}]. 

To determine the temperature and field dependencies of the critical currents and creep rates, magnetization studies were performed using Quantum Design SQUID magnetometers (fields up to $\SI{7}{\tesla}$) at Los Alamos National Lab and the Colorado School of Mines, and a Vibrating-sample Magnetometer (VSM) at the National High Magnetic Field Lab (fields up to $\SI{35}{\tesla}$). In all cases, the magnetic field was applied parallel to the c-axis (perpendicular to the film plane).  For the high-field VSM measurements, to observe measurably large signals over broad field ranges, we measured a stack of two identical cuts of each sample, effectively doubling the thickness and magnetization.

Vortex creep rates were determined by capturing the time-dependent magnetization $M(t)$ using standard methods,\cite{Yeshurun1996} which is expected to decay quasi-logarithmically over time $t$ as per Eq. \ref{eq:Jt}.  To measure creep, first, the critical state is established by sweeping the field $\Delta H>4H^\star$, where $H^\star$ is the minimum field at which magnetic flux fully penetrates the sample. Second, the field is set to the value of interest for the measurement and $M$ is then recorded every \textasciitilde $\SI{15}{\second}$ for an hour. We then subtract the background produced by the substrate and adjust the time to account for the difference between the initial application of the field and first measurement. This difference is found by maximizing the correlation coefficient of a linear fit to $\ln M - \ln t$.  Finally, the creep rate $S$ is extracted from the slope of the linear fit, per Eq. \ref{eq:Sdefinition}.

The critical current $J_c(T,H)$ was extracted from the magnetization data using the extended Bean critical-state model for rectangular samples.\cite{Bean1964a,Gyorgy1989a}
\begin{equation}
    J_c (T,H)=\frac{20\Delta M(T,H)}{h(1-w/3l)}
\end{equation}  Here, $\Delta M$ is the difference between the upper and lower branches of the magnetization loops $M(H)$, $h$ is the film thickness, $w = 3\text{-}\SI{5}{\milli\meter}$ and $l = 3\text{-}\SI{5}{\milli\meter}$ specify sample width and length.

\section{Results}

\subsection{Impact of film thickness: Elastic versus rigid vortex dynamics}

Beyond the bulk properties, it is important to recognize that the critical current $J_c$ and creep rate $S$ acquire an explicit dependence on the sample thickness $h$, once the latter drops below the pinning length $L_c$, i.e.\ $h \!<\! L_c$.  Phenomenologically, this represents a transition between elastic dynamics in thick (3D, bulk) samples and faster rigid vortex dynamics in thinner (quasi-2D) samples. That is, when $h > L_c$, $J_c$ and $S$ do not depend on thickness, whereas when $h<L_c$, both $S$ and $J_c$ become anticorrelated with thickness. Moreover, because the pinning length is temperature-dependent, so is the regime.

\begin{figure}[h!]
\begin{center}
    \includegraphics[width=0.85\linewidth]{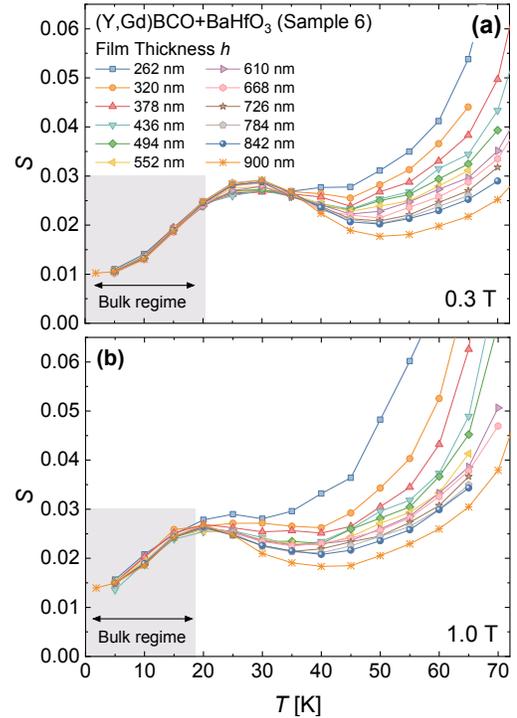}
    \caption{Thickness dependence of the vortex creep rate in sample 6, which contains a high density of BaHfO$_3$ nanoparticles, in fields of (a) \SI{0.3}{\tesla} and (b) \SI{1}{\tesla}. The creep rate is thickness-independent below $T \approx \SI{20}{\kelvin}$. \label{fig:STthickness}}
\end{center}
\end{figure}

\begin{figure*}[t]
    \centering
    \includegraphics[width=0.7\textwidth]{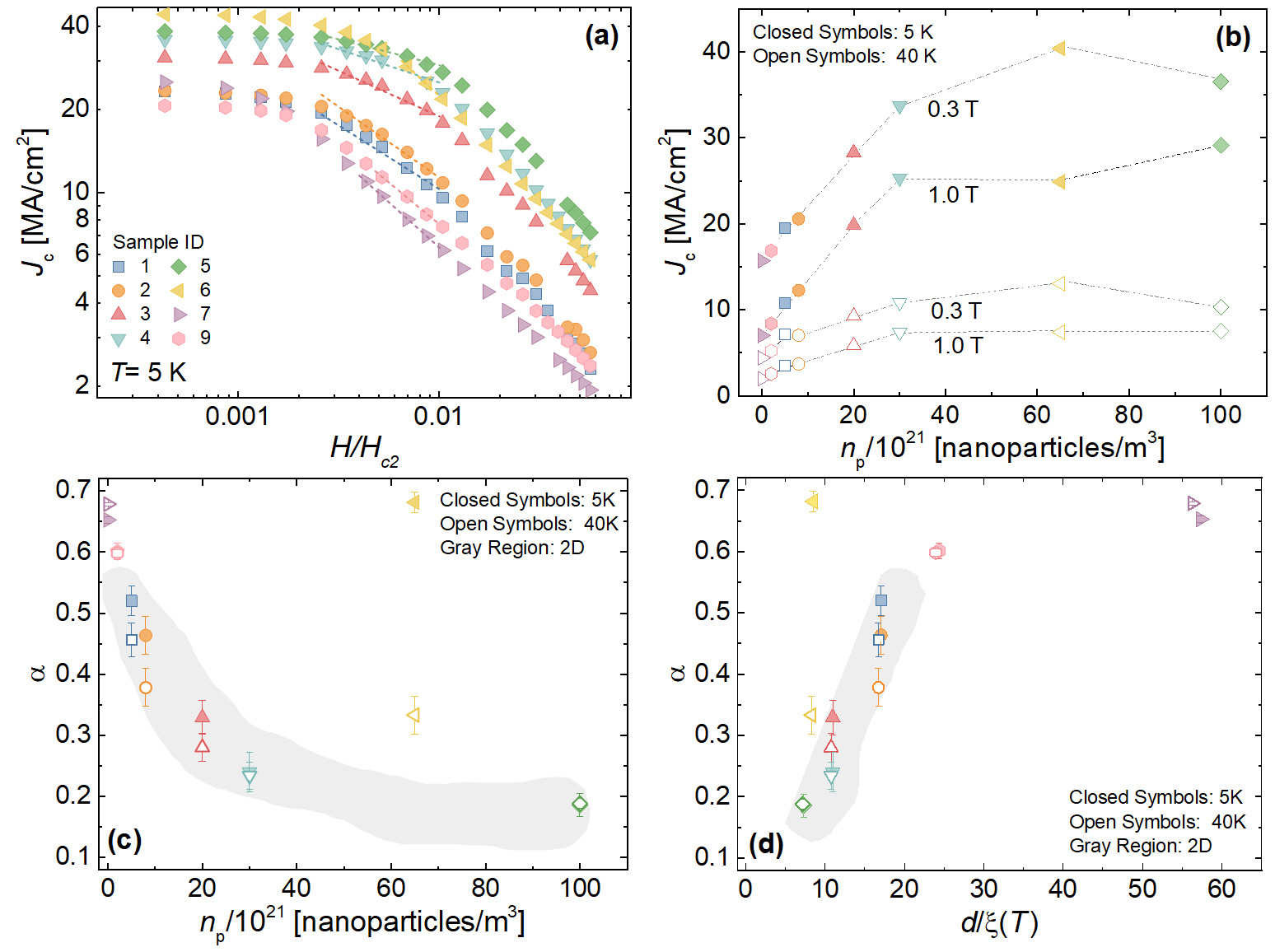}
    \caption{(a) Dependence of the critical current $J_c$ at $T = \SI{5}{\kelvin}$ on the magnetic field, normalized to the upper critical field $H_{c2}$. Dashed lines show region over which exponent $\alpha$ was extracted (between $H_{sf}/H_{c2}$ and 0.01). (b) $J_c$ versus nanoparticle density $n_p$ at $\SI{0.3}{\tesla}$ and $\SI{1}{\tesla}$. Dependence of $\alpha$ on (c) nanoparticle density and (d) nanoparticle diameter $d$, normalized to the coherence length. In (c) and (d), the gray regions highlight the data for 2D samples.}
    \label{fig:jc-and-alpha}
\end{figure*}

When the superconductor contains strong pinning centers in addition to point defects, this pinning length deviates from the predictions of weak collective pinning theory, e.g. $L_c(T)\approx(\xi/\gamma)(J_d/J_c)^{1/2}$ for single vortex dynamics, to a mixed pinning length $L_c^{mp} > L_c$ that can be extracted from creep measurements.\cite{Eley2018a} From Fig. \ref{fig:STthickness}, we expect bulk behavior in films of all tested thicknesses at $T=\SI{5}{\kelvin}$.

Our study contains three samples in the bulk-regime at all temperatures and five that host quasi-2D vortex dynamics at higher temperatures $T \gtrsim \SI{20}{\kelvin}$.  The thickness metric for categorizing the films was based off of a previous study\cite{Eley2018a} in which we extracted an effective pinning length $L_c$ from creep measurements collected on (Y,Gd)BCO films of a wide range of thicknesses as well as the data shown in Fig. \ref{fig:STthickness} for (Y,Gd)BCO films containing BaHfO$_3$ nanoparticles.  For clarity, throughout this work, we will analyze results from each category separately.

\subsection{Low-field critical current dependence on nanoparticle diameter and density}

Figure \ref{fig:jc-and-alpha}(a) shows the dependence of critical current $J_c$ on applied magnetic fields up to $\SI{7}{\tesla}$, normalized by the upper critical field $H_{c2}$, for all 8 samples. In Fig. \ref{fig:jc-and-alpha}(b), a snapshot of this data (at two fields and temperatures) is re-plotted versus the nanoparticle density, showing that increasing the density is indeed effective in boosting $J_c$, most dramatically at low temperature $T = \SI{5}{\kelvin}$. Focusing on the field dependence of $J_c$, we see that below the self-field $\mu_{0}H_{sf}=\mu_{0} \gamma J_{sf}h/\pi$, $J_c$ is field-independent, whereas above it, $J_c$ exhibits the expected power-law behavior $J_c \sim B^{-\alpha}$.  Here, the exponent $\alpha$ can provide information regarding the pinning regime. Consequently, the dependence of $\alpha$ on defect sizes and densities can reveal whether the microstructure induces weak collective or strong pinning and provide information on vortex-defect interactions, as discussed in Sec. \ref{sec:theory}. Hence, to understand the low-field $H \ll H_{c2}$ pinning regimes, we extract $\alpha$ from the slopes of linear fits to the data in Fig. \ref{fig:jc-and-alpha}(a) between the calculated $H_{sf}$ and $H/H_{c2} = 0.01$.  The extracted $\alpha$ parameters are plotted versus the nanoparticle density $n_p$ and normalized diameter $d/\xi$ in Figs. \ref{fig:jc-and-alpha}(c) and \ref{fig:jc-and-alpha}(d).

\begin{figure*}[ht!]
    \centering
    \includegraphics[width=1\textwidth]{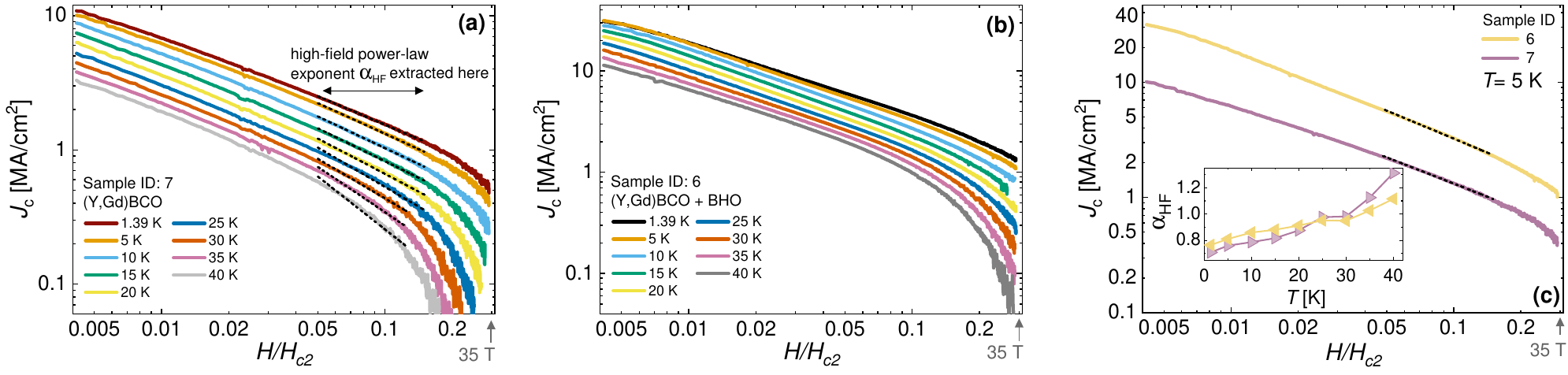}
    \caption{Field-dependent $J_c$ at various temperatures in (a) the reference sample and (b) sample 6, which contains BaHfO$_3$ nanoparticles and has the highest critical current of all samples in this study. Dashed lines in (a) show the regions over which $\alpha_{HF}$ values were extracted (0.05 < $H/H_{c2}$ < 0.15). (c) Comparison of $J_c$ in the reference sample and sample 6 at $T = \SI{5}{\kelvin}$. (Inset) Temperature dependence of extracted $\alpha_{HF}$. } \label{fig:Jc-alphas-highfield}
\end{figure*}

 Samples 6, 7 and 9 are all approximately $\SI{900}{\nano\meter}$ thick, therefore host three-dimensional vortex dynamics.~\cite{Eley2018a} For these samples, $\alpha$ ranges from 0.6 to 0.7 at \SI{5}{\kelvin}, which is within the expectations of strong pinning theory.~\cite{Willa2017} Additionally, within the measured magnetic field range, we find that $J_c$ is more sensitive to nanoparticle density than pinning volume (see Table \ref{tab:samples}) and many small defects are preferable to fewer large ones.

 In the thinner samples (\#1\text{-}\#5), the power-law exponent is more sensitive to the defect density and size than in the bulk samples. Figure \ref{fig:jc-and-alpha}(c) reveals a striking, systematic correlation between $\alpha$ and $n_p$, increasing from 0.2 at a high density of $100 \times 10^{21}/m^3$ to 0.53 for a lower density of $5 \times 10^{21}/m^3$, then only slightly increasing to \textasciitilde 0.6 for the reference sample containing only $0.03 \times 10^{21}/m^3$ 225 precipitates. This is remarkably consistent with the expectations from the TDGL simulations for low versus high nanoparticle densities. Empirically, we also observe a strong dependence of $\alpha$ on the defect size at nearly fixed volume fraction (e.g. comparing all films with a volume fraction of approximately 8.7-8.6\%). This result is in contrast with the weak dependence expected for three-dimensional strong pinning.\cite{Willa2017}

\subsection{Critical current at high magnetic fields}

To study strong pinning at high fields and look for the predicted peak effect, we measured samples 6 and 7 at the NHMFL in static fields up to $\SI{35}{\tesla}$ ($H/H_{c2} \approx 0.3$) using a VSM.  Accordingly, magnetization loops were collected for each sample at temperatures up to $\SI{40}{\kelvin}$, from which we calculated $J_c(H)$, shown in Figs. \ref{fig:Jc-alphas-highfield}(a) and (b) for samples 7 and 6, respectively. Figure \ref{fig:Jc-alphas-highfield}(c) shows how the enhancement in $J_c$ that is achieved by incorporating a high density of BaHfO$_3$ nanoparticles (sample 6) is maintained at high fields.

We extracted the high-field $\alpha$ (denoted $\alpha_{HF}$) from linear fits to $\log J_c$ versus $\log H/H_{c2}$, restricted to $H/H_{c2}=0.05 \text{-} 0.15$, for direct comparison with the strong pinning result in Ref.~[\onlinecite{Willa2017}].  For data collected at $T = 30\text{-}\SI{40}{\kelvin}$, the upper bound of the fit was decreased to $H/H_{c2}$ = 0.125 due to noise in the system as the signal approached its sensitivity limit. Figure \ref{fig:Jc-alphas-highfield}(c) compares results for the two samples and the inset displays the dependence of $\alpha$ on temperature.   From these comparisons, $\alpha$ appears independent of the nanoparticle diameter or density at these high magnetic fields. Furthermore, $\alpha \sim 0.8$ at \SI{5}{\kelvin} and $\alpha \sim 1.2\text{-}1.4$ at \SI{40}{\kelvin}.  This is consistent with the expectations of increased $\alpha \lesssim 1$ at high fields as each pinning site is occupied by a vortex.

According to strong pinning theory, $J_c(H)$ should exhibit non-monotonic behavior at high fields as nanoparticles capture multiple vortices. For sample 6, we would expect the lowest field peak to occur around $B \!\sim\! \Phi_{0} / d^{2} \approx \SI{10}{\tesla}$ and additional peaks at higher fields as the nanoparticle pinning site occupancy increases. However, we observe no non-monotonic behavior over the entire field range.  A few reasons may account for this discrepancy.  First, the distribution of nanoparticles in our samples may not be as homogeneous as represented in the simulations and the size variations noted in Fig. \ref{fig:TEM}(c) may smear the peak. In fact, variations in the diameter enter quadratically into the expected peak position. Second, all samples contain a sparse distribution of large 225 nanoparticle precipitates that likely capture multiple vortices at lower fields than the smaller Ba$M$O$_3$ inclusions.  However, the density of 225 precipitates may be too small for multiple vortex occupancy of these defects to induce a measurable change in the $J_c(H)$ dependencies. Furthermore, the presence of twin boundaries may effectively increase the field at which the multiple vortices occupy the Ba$M$O$_3$ inclusions to beyond our measurement range or further smear the expected peak. There is not yet a formalism that can consider mixed pinning landscapes --- microstructures that consists of multiple different types of defects. This is partially because it is notably complicated to concurrently consider defects of varying dimensionality: point-like (e.g. vacancies), 2D planar (e.g. twin boundaries, stacking faults), and 3D (e.g. nanoparticles).  Third, note that at higher temperatures, $J_c$ curves decreased to below the sensitivity of the VSM at fields around $\SI{25}{\tesla}$ such that we were unable to resolve $J_c$ beyond $H/H_{c2} \approx 0.2$.

\begin{figure*}[t!]
\centering
\includegraphics[width=0.7\textwidth]{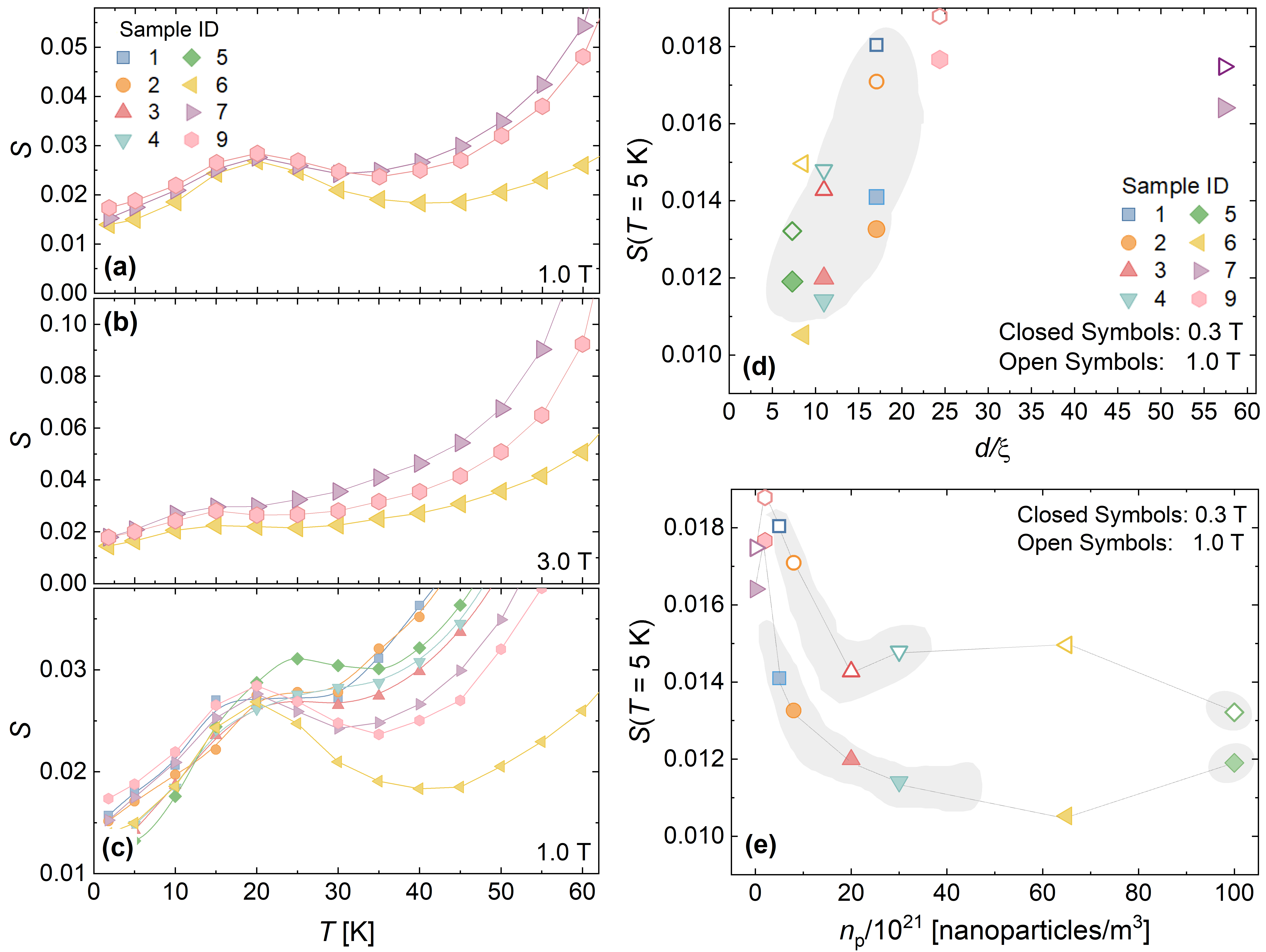}
  \caption{Direct comparisons of temperature-dependent creep in (a, b) bulk samples (\#6, 7, 9) at (a) 1.0 T and (b) 3.0 T, and (c) all samples at 1.0 T.  Creep at 5 K versus nanoparticle (d) size and (e) density.}
  \label{fig:creep}
\end{figure*}

\subsection{Vortex creep dependence on nanoparticle diameter and density}

\paragraph*{Bulk samples.} Figure \ref{fig:creep}(a,b) shows temperature-dependent creep rates for the three bulk films.  At \SI{1}{\tesla}, we observe the typical non-monotonic dependence reported in multiple studies of YBCO films containing nanoparticle precipitates.~\cite{Jia2013, Leroux2015, Miura2011b, Haberkorn2012d} In agreement with these earlier studies, see e.g.\ Ref.\ \sbonlinecite{Eley2017a}, we find that the non-monotonicity tends to disappear upon increasing the magnetic field.  The position of the peak in $S$ is the same in all samples, therefore, may depend on a common property, such as the density of 225 precipitates.  Additionally, from Figs. \ref{fig:creep}(a) and \ref{fig:creep}(b), we see that the Ba$M$O$_3$ nanoparticles have far less effect on the creep rate at low temperatures than at high temperatures, consistent with Ref.~[\onlinecite{Haberkorn2012d}]'s finding that defects generally affect creep more significantly at high-$T$.

At all fields, the film containing a high density of BaHfO$_3$ nanoparticles (sample 6) produces the lowest creep rates.  The film containing BaSnO$_3$ (sample 9) and the reference film (sample 7) produce similar creep rates at $\SI{1}{\tesla}$, however, at $\SI{3}{\tesla}$, creep in sample 9 is notably slower than in sample 7.  This suggests that the vortex density now is high enough that there is an insufficient number of 225 precipitates to pin all vortices and sample 9 benefits from the additional BaSnO$_3$ nanoparticles.

\paragraph*{All samples.} Figure \ref{fig:creep}(c) presents a comparison of creep in all samples. As expected based on results in Ref.~[\onlinecite{Eley2018a}], the thinner samples (1\text{-}5) exhibit little-to-no peak and faster creep than the bulk samples. Despite this difference, $S(T)$ at low temperatures is strikingly similar for all samples. Notably, creep increases approximately linearly with $T$, qualitatively adhering to the expectations of the Anderson-Kim model describing creep of rigid vortices $S \sim k_BT/U_c$.\cite{Anderson1964}  Moreover, all curves  deviate from this linear behavior around the same temperature $T = \SI{20}{\kelvin}$. Consequently, we can again see that that the effects of the inclusions are more significant at high temperatures than at low temperatures.

To view the dependence of the creep rate on nanoparticle properties and determine whether nanoparticles have an effect at low $T$, we plotted creep at $\SI{5}{\kelvin}$ versus the nanoparticle size and density in Fig. \ref{fig:creep}(d,e). A direct comparison of Figs. \ref{fig:creep}(d,e) and \ref{fig:jc-and-alpha}(c,d) reveals that, qualitatively, the dependence of $S$ on the defect density and size is very similar to that observed for the exponent $\alpha$. Specifically, creep shows a somewhat linear dependence on $d/\xi$ for all samples containing added nanoparticle inclusions at both \SI{0.3}{\tesla} and \SI{1.0}{\tesla}.  Data for the reference sample deviates from this trend, likely due to the large diameter of the dominate pinning sites---the 225 precipitates.  We also observe a decrease in creep values as nanoparticle density decreases.  Similar to the results for $\alpha(n_p, d/\xi)$ presented in Fig. \ref{fig:jc-and-alpha}(c), this decrease in the vortex creep rate suggests that a higher density of smaller defects may be preferable to a lower density of larger ones.

\subsubsection{Low-temperature creep rates}

It is common to compare creep data piecemeal to Eq. \ref{eq:ST}, such that at low temperatures $T \ll U_c/[k_B\ln(t/t_0)]$, the creep rate is expected to increase approximately linearly with temperature as $S \approx k_BT/U_c$.  Consequently, signatures of quantum creep are typically identified as a linear fit of $S$ that extrapolates out to finite values at zero temperature $S(T=0)>0$ or a creep rate that deviates from linear behavior at low temperatures. However, Eq.~\ref{eq:creepstrongpinning} shows that in the strong pinning scenario, thermal creep should cause a nonlinear, convex increase in $S$ at low temperatures, which becomes concave at the inflection point $T^\star = U_c/[5^{3/2}\ln(t/t_0)]$. \cite{Buchacek2019} Furthermore, a linear extrapolation at this inflection point indeed results in finite $S_0 = S(T=0)$ solely from thermal considerations.\cite{Buchacek2019} Hence, the extrapolation alone is insufficient evidence of quantum creep, without measurements in the quantum-regime.

\begin{figure}[h!]
\hspace*{-0.5cm}
\includegraphics[width=0.9\linewidth]{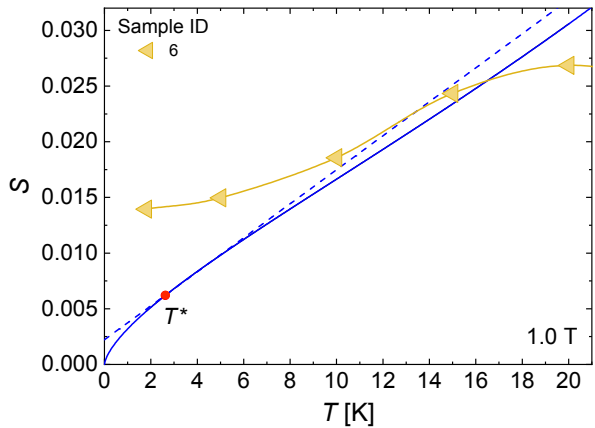}
\caption{(Yellow triangles) Low-temperature creep rate in a field of \SI{1}{\tesla} for sample 6.   Blue solid curve is a fit of the data to the expectations for thermal creep from strong pinning theory (Eq. \ref{eq:creepstrongpinning}) using $t_0 = 10^{-6}$. Deviations between experimental data and model possibly due to contributions from quantum creep. Dashed blue curve is a linear fit $k_BT/U_c$ around the inflection point $T^\star$. Notice that the dashed curve extrapolates to a finite $S>0$ as $T \rightarrow 0$, such that the conventional fit to a thermal region alone is insufficient evidence of quantum creep within the strong pinning paradigm.}
\label{fig:creepstrongpinning}
\end{figure}

In Fig. \ref{fig:creepstrongpinning}, we plot a fit of our creep data for sample 6 to Eq.~\ref{eq:creepstrongpinning}, where $U_c$ is the only free parameter and  $t_0 \sim 10^{-4}/\nu$ (for $H_{c2}/H \sim 100$ and $\nu \sim 10^{2}$ is related to the vortex velocity). From this fit, we extract $U_c/k_B \approx \SI{655}{\kelvin}$ and $T^\star \approx \SI{2.7}{\kelvin}$ using $t=\SI{3600}{\second}$ and demonstrate in the figure that a linear extrapolation around the inflection point leads to finite creep at zero temperature.  Note that our data does not only deviate from linearity at low $T$, but also from the functional form predicted for thermal creep from strong pinning centers, likely due to a significant, unknown contribution from quantum creep within our measurement range.

\subsubsection{Effect of Twin Boundaries}

Though we have focused on the effect of nanoparticles, it is important to mention that twin boundaries also likely contribute to slowing creep. Understanding the role of twin boundaries, planar defects prevalent in many crystalline materials, is especially challenging --– twin boundaries can have diametric roles in different regimes, either channeling or pinning vortices,\cite{Oussena1995a, Rouco2014a,Palau2006a} and the density of twin boundaries in films is difficult to control.   Twin boundaries act as strong pinning centers (pinning a higher density of vortices than the bulk), perturb the vortex lattice, and form a barrier for vortex crossing such that vortices will accumulate on one side of the twin boundary. \cite{Vlasko-Vlasov1994, Maggio-Aprile1997a} A better understanding of the effects of twin boundaries on vortex motion is required to further improve the performance of superconductors.

As vortices are not rigid objects, vortex motion will not simply be either arrested or funneled by a twin boundary.  The elasticity of a vortex means that thermal fluctuations can cause segments of the vortex to depin and form curved, half-loops that either slowly grow or rapidly expand if partially pinned by a neighboring defect.\cite{Sonin1995, Blatter1994, Blatter1991}  These mechanisms for vortex motion will each have different effects on $S$.

To better understand this, let us consider the temperature dependence of the twin boundary pinning potential.  Vortex attraction to twin boundaries is caused by suppression of the superconducting order parameter within the twin boundary, $\Psi_{TB}$.  This suppression extends a distance $\sqrt{2}\xi(T)=\sqrt{2}\xi_0{(1-T/T_c)}^{-1/2}$ around the twin boundary, where $\xi_0 \equiv \xi(0)$, such that the breadth increases greatly as $T~\rightarrow~T_c$.  The magnitude of this suppression is $\delta \Psi_{TB}=1-|\Psi_{TB}/\Psi_{bulk}|^2$, where $\Psi_{bulk}$ is the order parameter in the bulk infinitely far from the twin boundary.  While the exact temperature dependence of $\delta \Psi_{TB}$ is unknown, suppression of the order parameter is weak ($\delta \Psi_{TB}~\ll~1$) for $T~\ll~T_c \sqrt{1-[a /\xi (0)]^2)}\equiv AT_c$ and strong ($\delta \Psi_{TB}\sim 1-(\xi_0/a)^2[1-(T/T_c)^2] \lesssim 1$) in the opposite case,\cite{Blatter1994} where $a=\SI{3.8}{\angstrom}$ is the lattice parameter.  In YBCO samples,\cite{Blatter1994} $A\approx 0.97$ so the suppression is mostly nearly temperature independent, but should become stronger and temperature dependent as we approach $T_c$.  This increased effectiveness of twin boundaries in pinning at high temperatures is consistent with the slowest high-$T$ creep rates in the sample with the highest density of twin boundaries, and suggest that twin boundaries may contribute to lowering $S$ at high temperatures.

\section{Conclusion}
In conclusion, we conducted a thorough study of the effects of nanoparticle size and density on the critical current and vortex creep rates in (Y,Gd)BCO films. By conducting measurements in a wide range of temperatures $5\text{-}\SI{70}{\kelvin}$ and fields up to $\SI{35}{\tesla}$, we compared trends to the expectations of recent results from strong pinning theory. Though the $J_c\propto B^{-\alpha}$ dependence has been noted in numerous previous studies, we systematically unveiled a clear and similar dependence of $\alpha$ and $S$ on both the nanoparticle density $n_p$ and normalized size $d/\xi$, which is our main result.  Our results are consistent recent advances in strong pinning theory, which predicts a decrease in alpha from 0.66 to 0.2 with increasing $n_p$ and we observed a decrease from 0.65 to 0.2.  Lowering $\alpha$ signifies higher retention of $J_c$ upon exposure to magnetic fields, therefore, is favorable for applications.
At higher fields, \SIrange[range-phrase=\text{-}, range-units=single]{10}{35}{\tesla}, we observed larger $\alpha$ parameters ($0.8 \text{-} 1.4$), consistent with pinning sites becoming fully occupied. Lastly, the study showed a clear advantage to incorporating a high density of small nanoparticles versus a lower density of large ones.

\section*{Authors' contributions}
S.E. conceived and designed the experiment. S.E. and S.J. collected and analyzed the magnetization data. M.M. grew the samples, analyzed the microstructure, and assisted with data interpretation. R.Y. and T.K. performed transmission electron microscopy and other microstructural analysis. L.C. contributed to data interpretation. R.W. provided theory support and assisted with data interpretation. S.E., R.W., and S.J. wrote the manuscript. All author's commented on the manuscript.

\begin{acknowledgments}
We would like to thank Dr. Eun Sang Choi at NHMFL for assistance with high magnetic field measurements and Dr. Barry Zink and Mike Roos for use of their Quantum Design magnetometer to conduct preliminary measurements on stacked films in preparation for the NHMFL measurements. This material is based upon work supported by the National Science Foundation under Grant No. 1905909 (S.E., S.J.) and the U.S. Department of Energy, Office of Science, Office of Basic Energy Sciences (L.C.). A portion of this work was performed at the National High Magnetic Field Laboratory, which is supported by National Science Foundation Cooperative Agreement No. DMR-1644779* and the State of Florida.
R.W. was supported by the Deutsche Forschungsgemeinschaft (DFG, German Research Foundation) TRR 288-422213477 Elasto-Q-Mat (project B01), and by the Heidelberger Akademie der Wissenschaften (WIN08).
The work at Seikei University was supported by JST FOREST (Grant Number JPMJFR202G, Japan). A part of this work is supported JSPS KAKENHI (18KK0414 and 20H02184), PMAC for Science Research Promotion Fund and NEDO.
\end{acknowledgments}

\section*{Data Availability Statement}
The data that support the findings of this study are available from the corresponding author upon reasonable request.
\bibliography{maintex.bib}

\end{document}